# An intermediate-level physics laboratory:
# A system of two coupled oscillators with low-cost accelerometers

Mary Lamont and Minjoon Kouh
Physics Department, Drew University, Madison, NJ 07940

## Abstract

We describe an intermediate-level physics experiment beyond the first-year, which uses versatile, low-cost accelerometers (Wiimotes for the Nintendo Wii gaming system) and scientific-computing software for numerical data analysis. It is designed to help students to develop better understanding of a system of coupled oscillators and Fourier transform.

## Introduction

A system of coupled oscillators vibrates at special frequencies, known as the normal modes [1]. This important topic is typically covered in an advanced mechanics course with the emphasis on the theory. However, today's low-cost sensor technologies and scientific-computing software allow us to supplement the theoretical lessons with experiments. We have employed a wireless accelerometer (a Wii remote controller for the Nintendo game system) to measure the accelerations from two coupled oscillators and performed numerical analysis to study the normal modes.

Wiimote is a small, handheld game controlled for Nintendo Wii game system. This device can measure acceleration along three orthogonal axes and broadcast the data via Bluetooth. The data can be captured and recorded by free software like GlovePIE (http://www.glovepie.org). There have been a number of recent reports on the innovative usages of the Wiimotes in various physics laboratory and demonstration projects, which would have been too expensive or complicated to implement in the past. For example, Wiimotes have been used in the contexts of both introductory and advanced physics problems: vector decomposition on the inclined plane, harmonic and circular motions, collision, and other motion-tracking experiments [2,3,4].

Using commonly available equipments (springs, airtrack, and aircarts) and Wiimotes (Figure 1), we have designed a lab experiment for a coupled oscillator system, which would be appropriate for an intermediate-level mechanics course. This experiment can not only supplement the theoretical coverage (summarized below) of this important topic, but also provide an opportunity to teach the usage of scientific-computing software for data analysis.

## Summary of the theory

As covered in most standard mechanics textbooks [1], a system of two coupled oscillators has two degrees of freedom and two normal modes, commonly referred as the symmetric and antisymmetric modes. The symmetric mode is when both masses are displaced from equilibrium equally in the same direction. The antisymmetric mode is when the masses are displaced in the opposite directions by the same amount from equilibrium. In general, the masses will oscillate at a combination of the two modes, whose relative amplitudes depend on the initial conditions and can be calculated, as shown below.

The system of two coupled oscillators can be represented by a set of coupled differential equations:

Eqn. (1)

$$0 = m_1\ddot{x}_1 + (k + k_{12})x_1 - k_{12}x_2$$

$$0 = m_2\ddot{x}_2 + (k + k_{12})x_2 - k_{12}x_1$$

where $x_n$ is the position of the mass $m_n$ from the equilibrium (n = 1 or 2) and $k$'s denote the spring constants. With a trial solution, $x_n(t) = B_n \exp(i\omega t)$, we obtain a set of linear equations for $B_n$, where the condition for non-trivial solutions (i.e., non-zero $B_n$) is that the following determinant must be equal to zero:

Eqn. (2)

$$\begin{vmatrix} k + k_{12} - m_1\omega^2 & -k_{12} \\ -k_{12} & k + k_{12} - m_2\omega^2 \end{vmatrix} = 0$$

Solving for ω yields

Eqn. (3)

$$\omega^2 = \frac{(k + k_{12})(m_2 + m_1) \pm \sqrt{\left(-(k + k_{12})(m_2 + m_1)\right)^2 - 4(m_1 m_2)((k + k_{12})^2 - k_{12}^2)}}{2m_1 m_2}$$

When $m_1 = m_2 = m$,

Eqn. (4)

$$\omega = \sqrt{\frac{(k + k_{12} \pm k_{12})}{m}}$$

This well-known result reveals the normal modes, or the two frequencies at which the masses would oscillate. Further calculations (i.e., setting t=0 with Eqn. 13.10 in [1]) shows that the initial conditions, the position of the two masses at t=0, can be expressed as a combination of the amplitudes of the normal modes:

Eqn. (5)

$$x_1(0) = D_1 + D_2$$

$$x_2(0) = -D_1 + D_2$$

where $D_n$ is the total magnitude of each normal mode (i.e., the sum of amplitudes of the positive and negative frequency components, $+\omega$ and $-\omega$).

**Results**

**(1) Frequencies of the normal modes**

We simultaneously released two masses, aircarts plus Wiimotes, from various initial conditions, allowed them to oscillate freely for approximately 8 minutes until they stopped, and measured the accelerations. Three sample trials are shown in Figure 2, where one mass was displaced from -0.2 m from its equilibrium position, and the other mass was initially displaced at -0.2, 0.2, and 0 m (symmetric, antisymmetric, and mixed modes, respectively).

The fixed parameters in the experiments were:

| $m_1$ and $m_2$ | 0.50 kg |
|---|---|
| k (both) | 2.27 N/m |
| $k_{12}$ | 2.13 N/m |

We have performed the discrete Fourier analysis on the Wiimote data, using the FFT function in Matlab (Mathworks, Natick, MA), because according to the Fourier analysis, a general function can be represented as a sum of trigonometric functions, and, hence, our accelerometer signals would be approximated by a sum of two trigonometric components, corresponding to the symmetric and antisymmetric modes [5, 6, 7]. As expected and as shown in Figure 2, the Fourier transformation of a pure mode (either symmetric or antisymmetric) reveals a single peak, and that of a mixed mode reveals two peaks at the expected frequencies.

The full experiment was consisted of two initial conditions for one mass (-0.1 and -0.2 m) and seven conditions for the other (-0.3, -0.2, -0.1, 0.0, 0.1, 0.2 and 0.3 m). The frequencies of the two normal modes, as revealed by the Fourier analysis, were within 3 percent of the expected values.

| | Measured (Hz) | Expected (Hz) |
|---|---|---|
| $\omega_1$ (Symmetric) | $0.33 \pm 0.00$ | 0.34 |
| $\omega_2$ (Antiymmetric) | $0.57 \pm 0.01$ | 0.58 |

**(2) Relative amplitudes of the normal modes**

Unlike the frequencies, the relative amplitudes of the normal modes depend on the initial condition, as shown in Eqn. 5. We estimated the amplitude of each normal mode by integrating the area under the FFT curves around the peak within the 0.1 Hz window. The ratios of the symmetric and antisymmetric modes were compared against the theoretical values of $D_1/D_2$, as shown in Figure 3(a). Note that the acceleration is the second derivative of the displacement ($\ddot{x} = -B\omega^2 \exp(i\omega t)$ and $x = B\exp(i\omega t)$.), so the amplitudes obtained from the Wiimote data have to be divided by the frequency squared, before being compared to the theoretical values obtained from the displacement variables in Eqn. 5.

**(3) Mass dependence**

In another set of experiments, additional masses, between 0.0 and 1.0 kg in 0.2 kg increment, were added to one of the carts. The cart with the additional mass was displaced -0.2 m from the equilibrium position, and the other cart started from its equilibrium. The Wiimote data was analyzed again with FFT, and the frequencies of the normal modes were found from the peaks of the Fourier analysis. As described by Eqn. 3, the added mass "slows down" the oscillation with its inertia, and the frequencies of the normal modes decreases, as shown in Figure 3(b).

**Discussion**

By incorporating low-cost and versatile accelerometers (such as the Wiimotes) and scientific-computing software (such as Matlab), we have designed an intermediate-level classical mechanics experiment, suitable for a typical sophomore or junior physics student, allowing them to study the important topic of coupled oscillators in quantitative details. Such experimental explorations not only deepen the understanding of the theoretical concepts of normal modes and Fourier analysis, but also provide an opportunity to learn computational tools and challenge the students to think carefully about the measurements and the theory (e.g., the difference between the accelerometer data and the results derived from position-based differential equations). With its $40 price, 100-Hz sampling rate, and within-5% accuracy, this relatively new sensor technology opens up many possibilities for innovative lab experiments beyond the first year of physics curriculum.

**Acknowledgement**

The authors would like to thank Stephen Takacs for his help with the setup of the experiment, Alae Kawam (for the idea of using two aircarts to give enough lifts) and Robert Murawski (for helping us to realize that the extra $\omega^2$ factors were missing in the amplitude analysis).

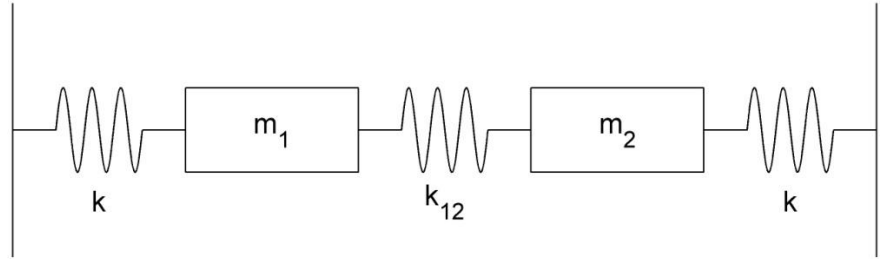

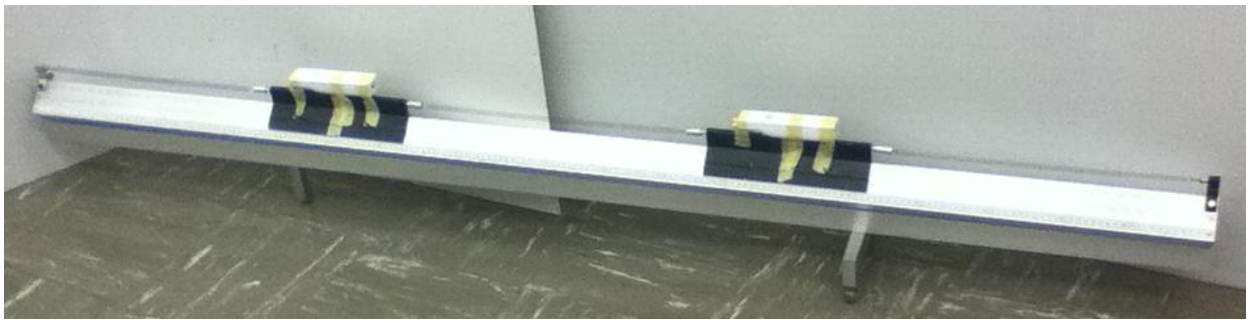

Figure 1. Experimental setup. Two aircarts per Wiimote were necessary to provide enough lift on the airtrack.

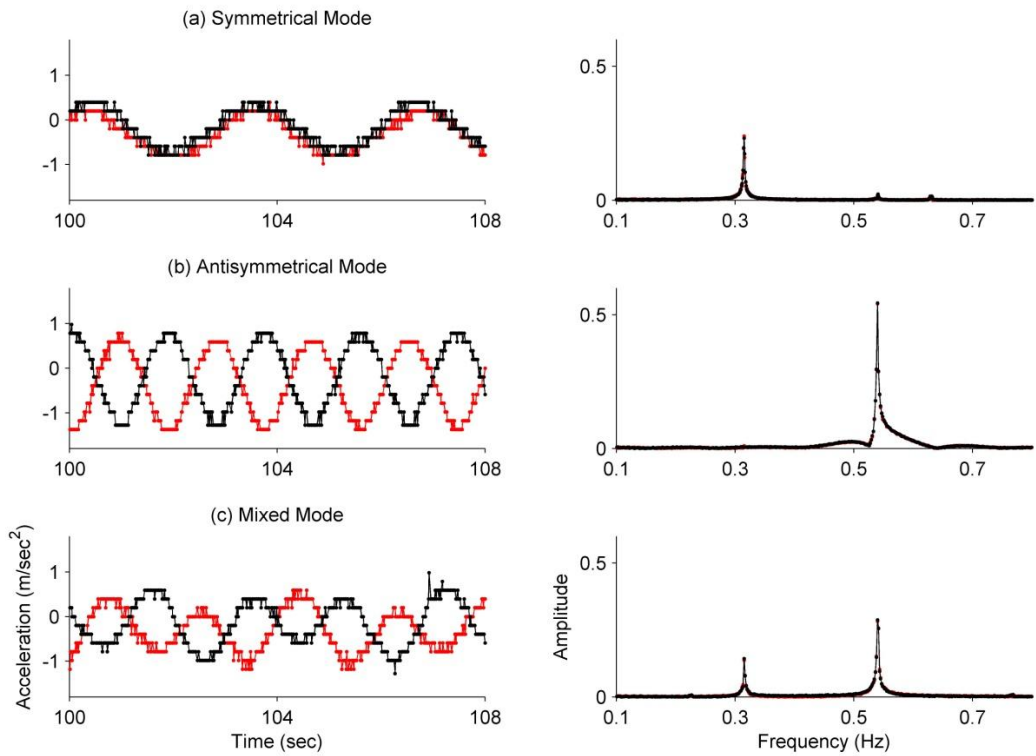

Figure 2. Sample trials with different initial positions of two aircarts. The accelerometer data from two Wiimotes (left) and their FFT analyses (right) are shown.

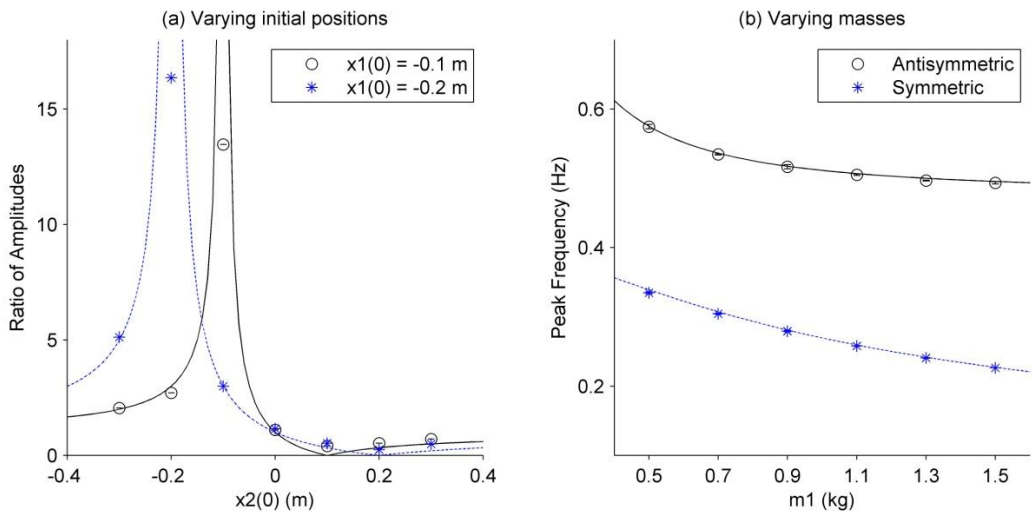

Figure 3. (a) The ratio of the amplitudes of the symmetric and antisymmetric modes at different initial conditions. The averages and standard deviations of three trials are shown. The amplitude of each normal mode was estimated by integrating the FFT amplitudes around its peak frequency within the 0.1 Hz window. (b) The frequencies of the normal modes with different amount of mass.